\pdfoutput=1

\documentclass[11pt]{article}

\usepackage[]{acl}
\usepackage{times}
\usepackage{latexsym}
\usepackage[T1]{fontenc}
\usepackage[utf8]{inputenc}
\usepackage{microtype}

\usepackage{subcaption}
\usepackage{graphicx}
\usepackage{appendix}
\usepackage{booktabs}
\usepackage{multirow}
\usepackage{tabularx}
\usepackage[normalem]{ulem}
\usepackage[capitalise]{cleveref}
\usepackage{xspace}
\usepackage{xspace}
\usepackage{multirow}
\usepackage{hyperref}
\usepackage{listings}
\usepackage{tabularx}

\title{Open Assistant Toolkit - version 2}

\author{Sophie Fischer$^{1}$ \And 
  Federico Rossetto$^{2}$ \And
  Carlos Gemmell$^{2}$\And
  Andrew Ramsay$^{2}$\AND
  Iain Mackie$^{2}$\And
  Philip Zubel$^{2}$\And
  Niklas Tecklenburg$^{2}$\And
  Jeffrey Dalton$^{1}$ \AND
  \textmd{$^{1}$University of Edinburgh, Edinburgh, United Kingdom} \\
  $^{2}$University of Glasgow, Glasgow, United Kingdom \\ 
  \texttt{s.fischer@ed.ac.uk} \\
}

\begin{document}

\newcommand{\OAT}{\texttt{OAT-v2}\xspace}
\newcommand{\code}[1]{\texttt{#1}}

\maketitle

\begin{abstract}
    We present the second version of the Open Assistant Toolkit (\OAT), an open-source task-oriented conversational system for composing generative neural models.
    \OAT is a scalable and flexible assistant platform supporting multiple domains and modalities of user interaction.
    It splits processing a user utterance into modular system components, including submodules such as action code generation, multimodal content retrieval, and knowledge-augmented response generation.
    Developed over multiple years of the Alexa TaskBot challenge, \OAT is a proven system that enables scalable and robust experimentation in experimental and real-world deployment. 
    \OAT provides open models and software for research and commercial applications to enable the future of multimodal virtual assistants across diverse applications and types of rich interaction.
\end{abstract}

\section{Introduction}

\begin{figure*}[tb]
    \centering
    \includegraphics[width=0.95\textwidth]{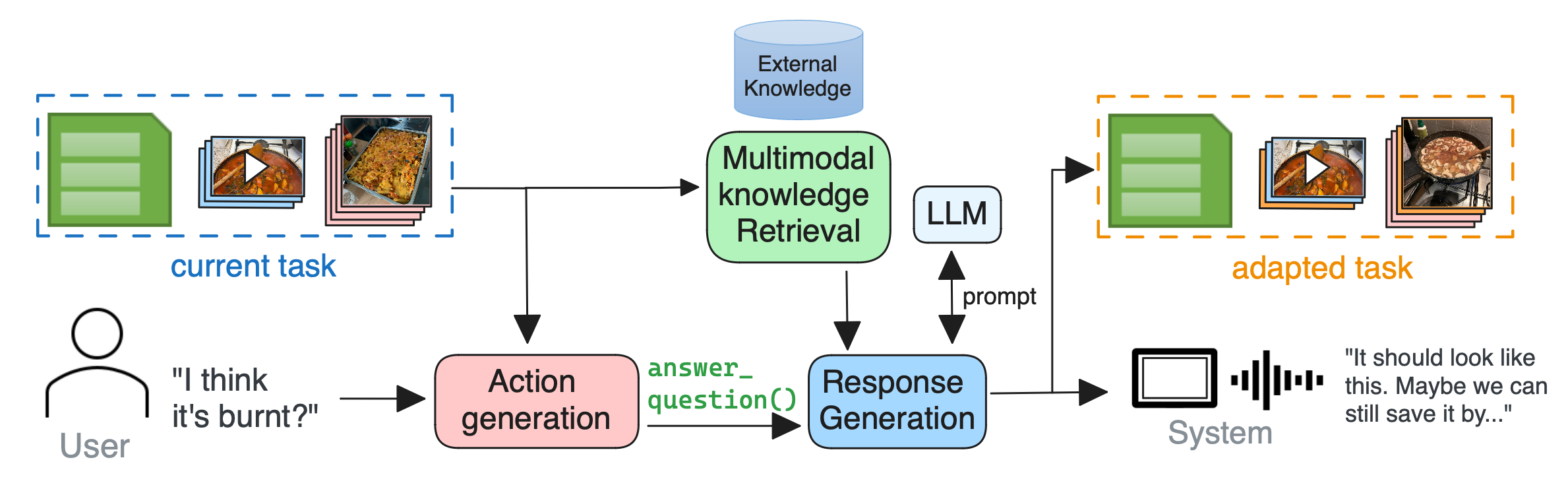}
    \caption{\OAT processes user utterances with different submodules. The system generates a system action based on the current task, system state and user utterance. Then, a knowledge-augmented generator generates a response based on the task and external knowledge. The system internally updates the task and responds to the user.}
    \label{fig:oat-end-to-end}
\end{figure*}

Building fluent and reliable conversational agents to assist with real-world tasks has been a focus of industry and academic world for many years.
Despite this fact, few modern open-source conversational frameworks are available. 

We therefore built the first version of the Open Assistant Toolkit (OAT v1) during the inaugural Alexa Prize TaskBot Challenge \citep{gottardi2022alexa} as the framework for the competition submission GRILLBot \citep{gemmell2022grillbot}.
OAT v1 is an open-source modular task-oriented conversational framework to create structured multimodal dialogue.

In this work, we present the new \OAT extension, which enables composing generative neural models by providing specialised models and LLM endpoints.
Key online components include action code generation, multimodal knowledge retrieval and response generation, which interact with the underlying task state.
The generative power of \OAT assistants allows knowledge-grounded question answering and adapting the task at hand live to suit user preferences \citep{fischer2024grillbot}. 

Starting as an alternative to Alexa, it now supports different front-end clients and comes with voice and multimodal integration to enable real-world task assistance.
To our knowledge, none of the few modern open-source conversational frameworks are task-oriented and can support multimodal and LLM integration \citep{Pydial_Ultes2017, GENIE_Campagna2019, zharikova2023deeppavlov}.

One requirement of a usable task assistant is reliability and in-depth domain knowledge. 
To source knowledge and tasks, we release a new offline pipeline to parse and augment task data from CommonCrawl.
The offline pipeline supports using LLMs and additional open-source multimodal data sources to enhance tasks and make them more suitable for real-world assistance.
In the online system, all domain knowledge is grounded with specific task indices and knowledge sources generated with the offline pipeline.

To ensure reliable answer times, \OAT uses a modular setup using Docker and Kubernetes.
The system is scalable, lightweight, and has a non-resource-intensive architecture with low latency.
In addition, the system is ready to use and deployable with newly released models and datasets.

To summarise, \OAT includes the following contributions:
\begin{itemize}
    \item New model training data and model releases.
    \item New conversational policies to handle the backend logic of extended domain-specific system actions.
    \item LLM infrastructure: Generative online LLM extension for seamless zero-shot prompting, task-specific knowledge-augmented question answering, guided search flows and themes.
    \item Offline tools: Offline pipeline to extract, generate and synthetically enhance tasks offline for engaging conversational content using LLMs and knowledge sources.
    \item Specialised Model training pipeline: Create and extend specialised LLMs for time-critical subtasks such as the NDP.
\end{itemize}

\OAT is available on GitHub\footnote{https://github.com/grill-lab/OAT}.

\section{\OAT}

\OAT was produced for the second Alexa Prize TaskBot Challenge \citep{agichtein2023alexa}.
Changes in \OAT reflect overall framework improvements made for GRILLBot-v2 \citep{fischergrillbot}.

Building on previous work, \cref{fig:oat-end-to-end} shows how \OAT splits action generation, textual response generation and knowledge retrieval into different modules.
\OAT generates system actions with a NDP-style model.
For answer generation, depending on the system action, \textit{Orchestrator} policies handle multimodal knowledge retrieval and generating of grounded and fluent responses.
For response generation, \OAT enables passing context and specific prompts into different LLMs using the Huggingface Text Generation Interface (TGI).
This enables appending dialogue, context and prompts to different models flexibly without fine-tuning and generating flexible and grounded responses.
We adapt the system to perform knowledge-augmented response generation using TGI, improve action generation with specialised models and add new multimodal retrieval models.
Furthermore, we add different search mechanisms and synthetic task generation to find more relevant tasks which suit the user's needs.

In addition to the online architecture, we also release the training and offline pipeline.
With the training pipeline, we release an interface to allow easy adaptation of specialised models via fine-tuning.
Specialised models have the advantage that they keep latency low.
An example is the NDP for system action generation or fine-tuned QA models.
We envision further extension to use specialised models for tool use, such as tabular QA from structured knowledge from external knowledge sources \citep{ToolWriter_Gemmell2023}.

The offline pipeline transforms human-written websites into executable TaskGraphs.
This ensures that retrieved tasks are based on real-world seeds written by humans.
This is critical for tasks that affect the real world, such as cooking and home improvement.

Online, offline and training components will allow researchers to adapt and advance OAT to their needs flexibly.

\subsection{Dockerised modular architecture}

\begin{figure}[tb]
    \centering
    \includegraphics[width=0.4\textwidth]{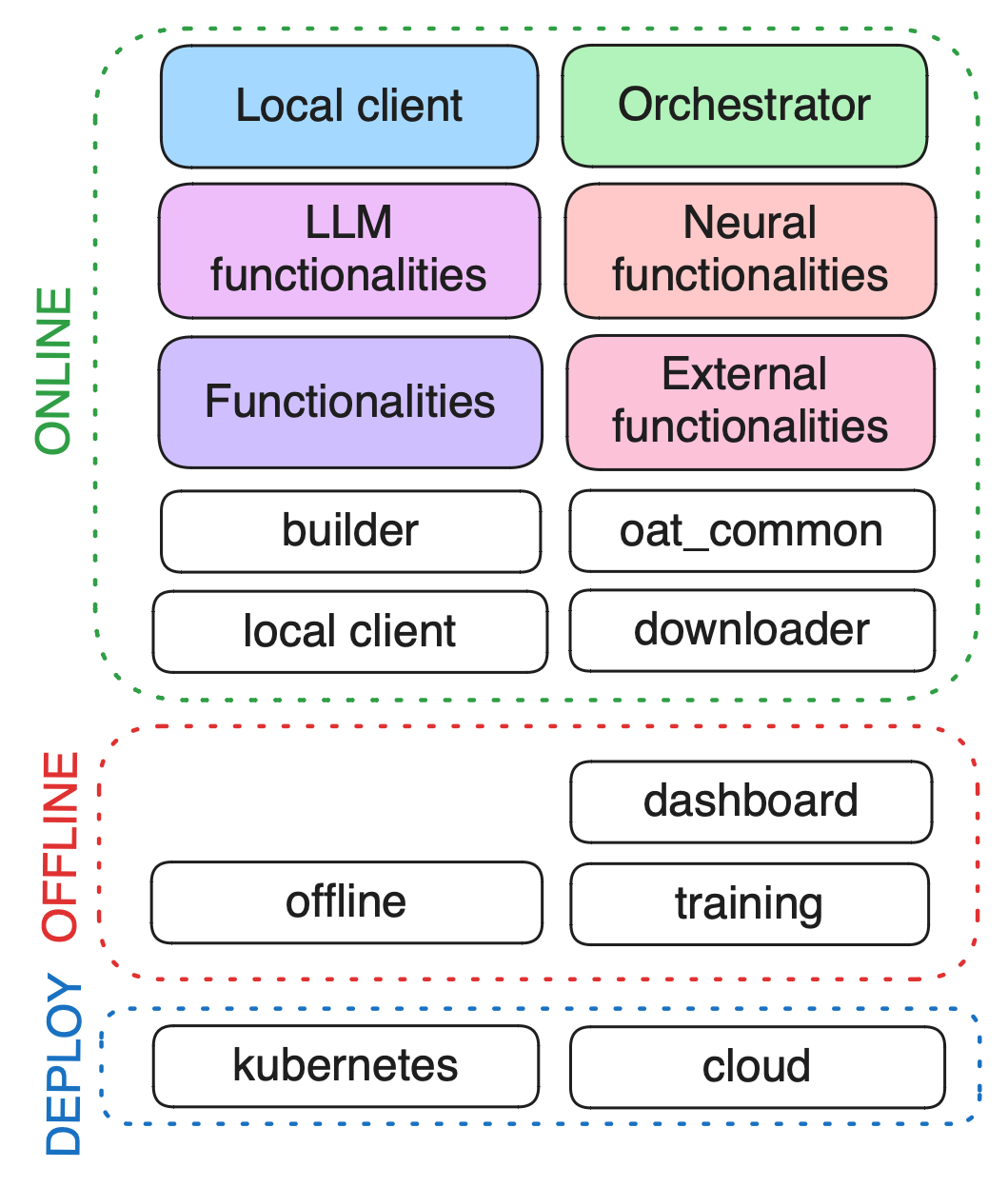}
    \caption{\OAT code base is structured into online, offline and deployment Docker containers. Coloured components will show up in the logs when running the online system.}
    \label{fig:docker-types}
\end{figure}

We implement \OAT in a modularised format to help with version control, decoupling and deployment.
\cref{fig:docker-types} shows all of \OAT Docker containers.
They are split into online, offline and deploy(ment) containers.
To run the online system, we use \textit{docker compose}, which spins up and connects all the services through environment variables.

The docker environment is an Ubuntu instance which runs both the client and server.
All Docker containers inherit from the base Docker image, defined in \textit{oat\_common}.
Within each Docker container, we use gRPC\footnote{https://grpc.io/} to define endpoint visible to other containers using protocol buffers.
The \textit{builder} builds the protobuf servers and objects.
All the required default files to run the system are downloaded by spinning the containers up by the downloader.

Beyond live system components, we provide additional offline features with OAT v1 for easy development and demos.
The offline Docker containers include the \textit{dashboard} used to review past conversations.
Furthermore, we release our offline and training containers, which can be used to create the new offline artefacts and train the NDP.

\section{Online Infrastructure}
In this section, we review additions to OAT made for version 2.
We review how the NDP generated action code generation for different domain adaptations.
We then introduce our LLM extension and components using it.
This includes the composed system's response generation using different system policies and live task adaptation.

\subsection{Code generation for dialogue management}

\begin{figure}[tb]
    \centering
    \includegraphics[width=0.45\textwidth]{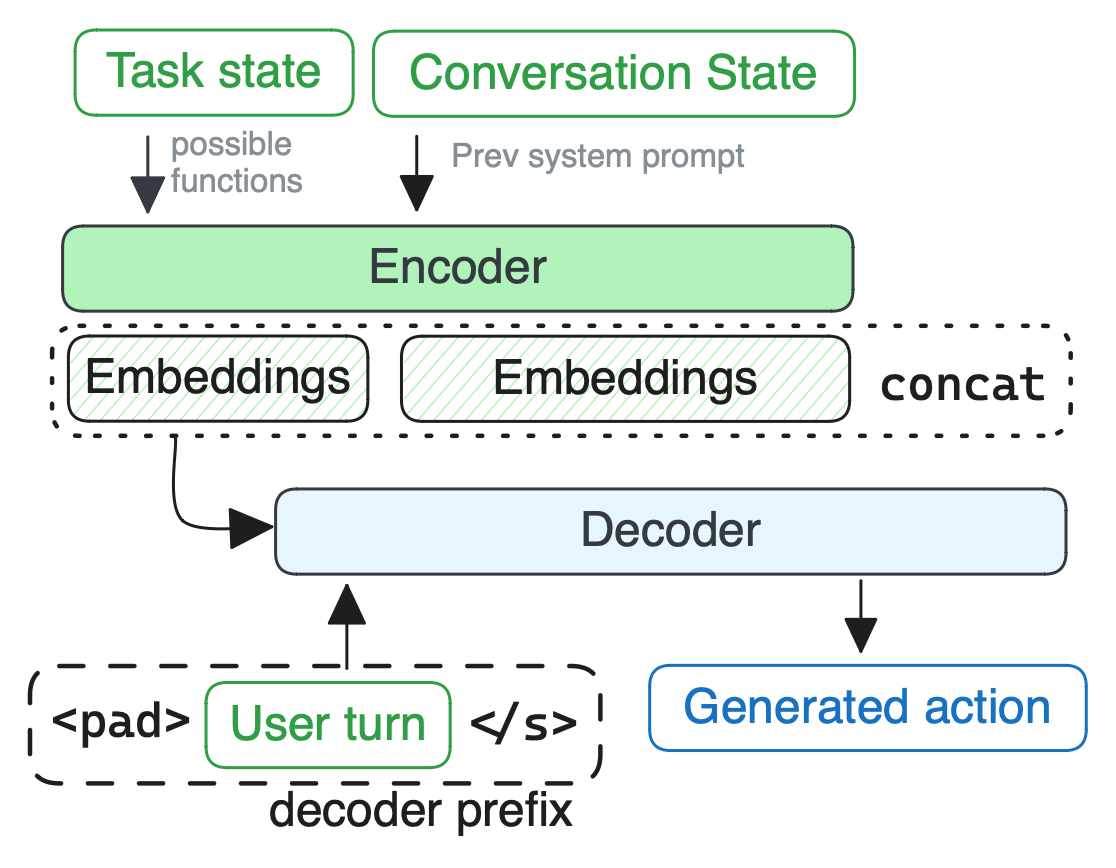}
    \caption{NDP architecture. The encoder embeds possible functions depending on the task state and the previous system prompt. The decoder prefix includes the user turn. The decoder then auto-regressively generates a system action in the action target space.}
    \label{fig:ndp_architecture}
\end{figure}

Based on previous work \citep{gemmell2022grillbot}, a Neural Decision Parser (NDP) model is \OAT's action generator.
It ingests a user utterance and the system state and generates code to represent the action the system should respond with.
\cref{fig:ndp_architecture} shows \OAT model's encoder-decoder architecture.

\begin{table}[tb]
\centering
\caption{Action space for NDP code generation. The action space is limited to avoid hallucinations, but the arguments can be flexibly generated. }
\label{tab:ndp-action-space}
\begin{tabularx}{0.5\textwidth}{lX}
\toprule
Action type                  & Specific Action                  \\ \midrule
Task navigation & restart, next, previous, select(int), step\_select(int), show\_more\_results                                                           \\
Conversational & ask\_question, search(query: string), show\_more\_details, chit\_chat   \\
General navigation           & start\_task, stop, pause, cancel, repeat, yes, no \\
Domain-specific              & set\_timer, confused\_user, show\_requirements, inform\_capabilities  \\ 
\bottomrule
\end{tabularx}
\end{table}

\begin{figure*}[tb]
    \centering
    \includegraphics[width=\textwidth]{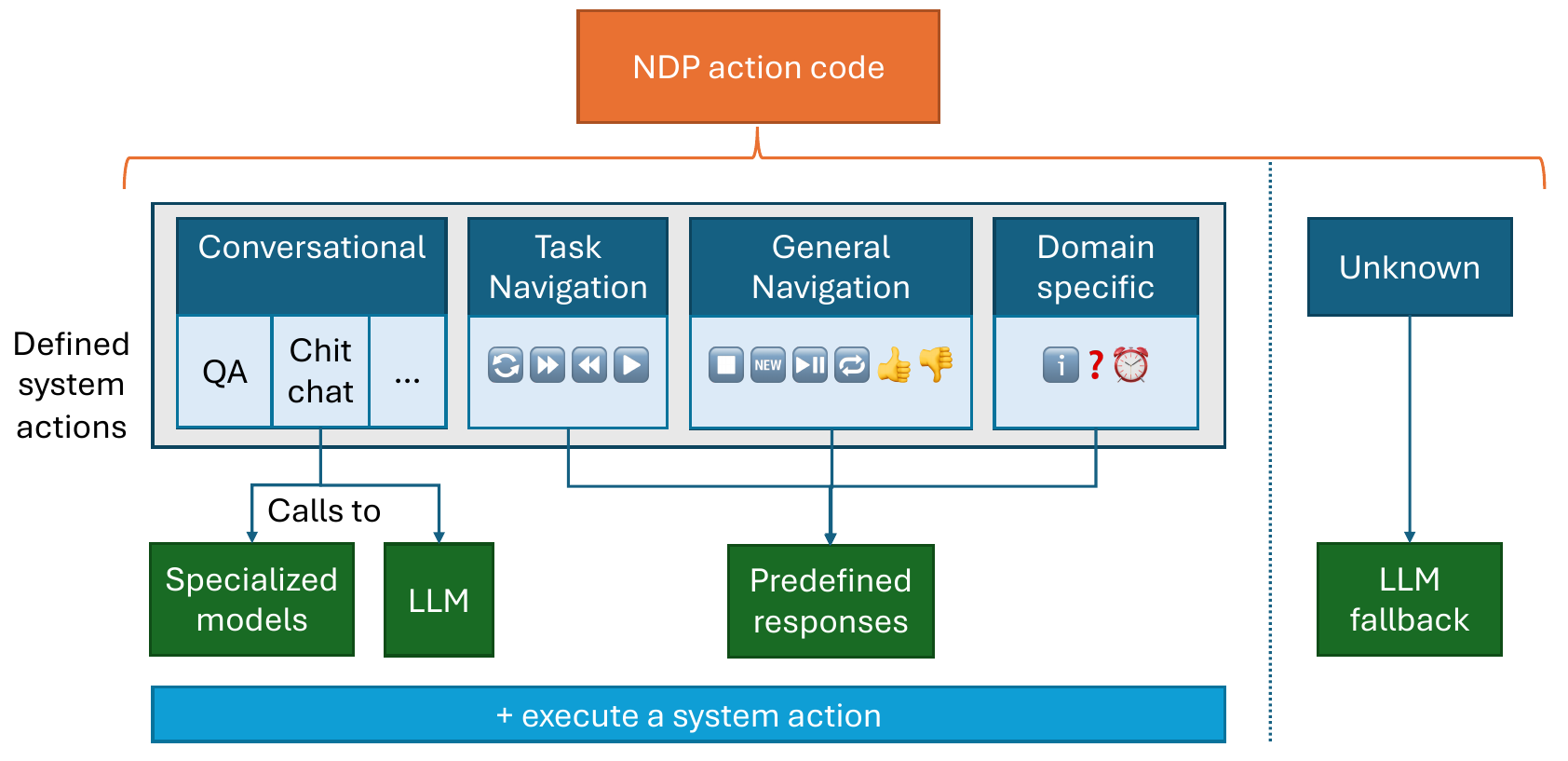}
    \caption{\OAT's composed system for response generation depends on NDP action generation. Depending on the action type, the \textit{Orchestrator} component handles calling different functionalities such as specialised models, LLMs, or predefined logic flows. If the NDP action code is unknown, a fallback LLM handles generating a fluent response and communicating system abilities to the user. }
    \label{fig:response_generation}
\end{figure*}

To create an NDP model, we define the action target space for all basic flows the model should support.
Each NDP code in the action space corresponds to a system action to modify the session and user state within the conversation.
\cref{tab:ndp-action-space} shows OAT's current NDP's action space.
The NDP model can also generate actions not in the action space, which a fallback LLM system handles by providing a fluent response.
No system action is executed in that case.

To add a new domain or conversation flows to \OAT, we need new tasks, new training data for the NDP and backend code to handle the domain-specific system actions.
For example, a chatbot assisting users with Finance supports different user tasks to a Cooking and DIY assistant such as GRILLBot \citep{fischergrillbot}.

To train, evaluate and test OAT's NDP, we use a synthetically generated and human-written dataset containing $\sim$1200 manually reviewed training data pairs.
We release the dataset and training pipeline in the \OAT release.
Training data pairs consist of a user utterance, the previous system response, the possible action target space and the annotated correct prediction.

\subsection{LLM Generation using TGI}
\OAT uses a locally deployable LLMs (e.g. \citet{Llama_Touvron2023, alpaca_stanford}) for zero-shot prompting during execution.
This allows fluent responding to dynamic user environments for question answering, live task adaptation or fallback chit-chat.

An LLM in a live deployed system needs to be scalable and easily interchangeable for experimentation.
GRILLBot-v2 uses locally deployed models, which are scalable within the framework but don't allow easy updating of weights and models.
Therefore, we extend the platform to interact with Huggingface's Text Generation Interface (TGI) \footnote{https://github.com/huggingface/text-generation-inference} to create standard input/ outputs for models.
Local models are deployed with predefined TGI Docker containers, which handle downloading, scaling, and keeping up-to-date with model versions without model-specific implementations.
The \textit{llm functionalities} container interacts with the deployed models and queries them using dialogue and context.

\begin{figure*}[tb]
    \centering
    \includegraphics[width=\textwidth]{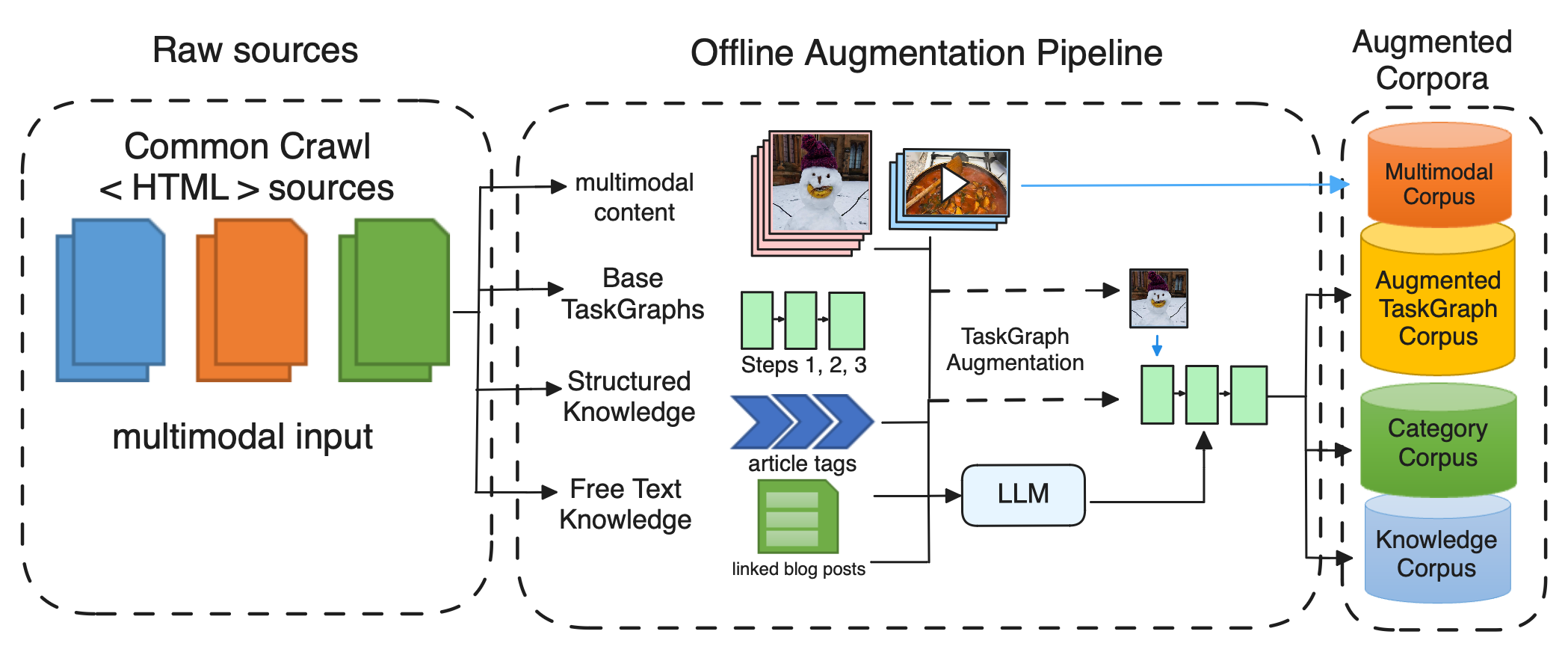}
    \caption{\OAT's offline pipeline ingests multimodal input from Common Crawl and parses them into subcomponents. Then, the pipeline transforms the data into augmented TaskGraphs. Other Corpora include categories and knowledge corpora and are written out for further online system use. }
    \label{fig:offline_pipeline}
\end{figure*}

\subsection{Composed response generation}
\cref{fig:response_generation} shows the current composed system's backend handling of the generated NDP actions.
We define backend logic to perform system actions for all defined actions.
Conversational NDP actions include question-answering, chit-chatting, asking for more details, requesting substitutions and searching.
We prompt a mix of specialised models and LLMs to run generation and knowledge retrieval.
We query self-hosted LLM endpoints with custom prompts and contexts depending on conversational state and question type.

For unseen NDP action codes, the system can't execute system actions since they are not defined in the backend handling within system functionalities.
Therefore, a LLM fallback model handles producing a fluent response.
All responses are post-processed and checked for safety.
We ensure limited hallucinations by prompting the model to include the system's capabilities.
During the deployment of GRILLBot, we saw 30\% of utterances before task selection fall into this category \citep{fischer2024grillbot}, highlighting the importance of providing a stable fallback answering system.

\subsection{Live task adaptation}
A task assistant in the real world encounters various changes to the underlying task due to the dynamic nature of user preferences and user environment.
For example, a user could not have an ingredient, run out of glue during home improvement or not like press-ups as part of their daily workout suggestion.
A key component to an engaging task-oriented conversation is the ability to adapt and modify a task based on the user's skill, available requirements and preferences.

Due to the TaskGraph representation, online services of \OAT can interact with the underlying task flexibly and modify task components live.
In previous work, we build support for performing live ingredient adaptation in the Cooking domain \citep{fischer2024grillbot}.

\OAT's online architecture supports calling LLM to modify the task due to the generated system action.
The system can reschedule, remove and append additional steps due to the TaskGraph representation.
TaskGraphs also enable new conversational paths, system initiative and linking multimodal content.

\section{Offline infrastructure}
As part of the offline pipeline, \OAT includes the previously unpublished offline pipeline to generate TaskGraphs.
We present our approaches for task augmentation and synthetic task generation.
Lastly, we release our training pipeline for the system's NDP.

\subsection{Offline pipeline}

\label{offline_pipeline}
With \OAT, we release our offline pipeline for generating offline artefacts.
\cref{fig:offline_pipeline} shows the run-able end-to-end highly configurable pipeline.
The pipeline supports extracting, parsing and augmenting publicly available multimodal data from Common Crawl into data components to use in the \OAT framework.
We release a set of parsers which extract the basic task data, multimodal components such as images and videos, structured knowledge found on webpages such as breadcrumbs and article tags, linked blog posts, and long text paragraphs as free text knowledge.

Since all tasks are written for online web interface usage, we add augmentation steps to transform tasks into engaging written and spoken walk-throughs.
Augmentations rewrite steps and add additional content to tasks using LLMs and external knowledge sources.
In previous work, we worked on aligning videos to task steps, which we implement as an example in the offline pipeline \citep{fischer2022vilt}.
We also support augment tasks to include images using contrastive loss \citep{CLIP_Radford2021, SBERT_Reimers2019}.

Furthermore, the new \OAT pipeline supports creating synthetic search trajectories by grouping thematically similar queries.
If the live NDP detects a vague search, those trajectories allow for eliciting preferences and guiding the user to a specific task.

\subsection{Synthetic task generation}

Using the Huggingface TGI extension, we can quickly spin up a self-contained LLM to call for task augmentation.
We create synthetic task components like additional details and step improvements and generate matching images \citep{Llama_Touvron2023, stable_diffusion} for the most asked user queries to ensure a relevant and engaging user experience.

We also experiment with fully synthetic task generations with larger models (e.g. \citep{ChatGPT}).
Related work showcases that whilst LLMs generate fluent task content, the generated tasks can be factually incorrect and potentially dangerous \citep{fischergrillbot, TWIZ2023}.
This shows the importance of having a retrieval corpora and a structured infrastructure to ensure task grounding and avoid potentially dangerous hallucinations. 
To ensure the factual correctness of recipes, we only use augmented tasks, not completely synthetically generated.

\subsection{Training pipeline}

In previous work, we compare the advantages and drawbacks of using LLMs versus specialised models for system components \citep{fischer2024grillbot}.
We release the training pipeline for the NDP to showcase how we train specialised models. 
Training data pairs are formatted correctly and passed to the encoder during pre-processing.
The encoder embeds the input, concatenates and formats it into batches to pass to the decoder.
The decoder then generates a system action based on the decoder prefix, which includes the embedded user turn.
This setup allows adapting the NDP with new training examples to new domains.

\section{Conclusion \& Future Work}

We present \OAT, a conversational framework composing neural generative models for conversational task assistants.
The release contains a deployment-ready online system and tools to generate offline artefacts, train component models, query LLMs and review conversational data.

Due to the rapid pace of LLM development in recent years, we envision \OAT as an interface for easy experimentation of grounded, deployment-ready, generative conversational task assistants.
Our vision is to allow transparent development of research systems without paywalls or closed-source models.
Integrating Huggingface's TGI LLM deployment module allows flexible experimentation with different models directly in a natural conversational setting.

We envision extending our work to include multimodal LLMs and further visual input into OAT in future work.
We envision leveraging vision language models to allow reasoning over retrieved relevant multimodal content.
This could include narrating and finding relevant sections in related videos.
These more complicated system operations require advanced integration of tool use to assist with shopping for required materials, bookings, and communication with IoT devices.
In addition, visual input to OAT could be done with Augmented Reality devices, such as Smart glasses or more advanced camera input via devices like Amazon Alexa.
The next generation of interactive assistants will be able to "look over your shoulder" to truly assist users with real-world tasks.

\clearpage
\bibliography{anthology,custom}

\clearpage

\end{document}